\newcommand{\h}[1]{\mathop{\lambda}\limits_{#1}\ \!\!\!}

\documentstyle{article}
\begin{document}
\begin{center}
{\huge
{\bf Lense-Thirring Field  and \newline
 the Solar Limb Effect}} \newline
\\

{\it \bf  M.I. Wanas$^{1,3}$, A.B. Morcos$^{2,3}$ and S.I. El Gammal$^2$}\newline
\\

  $^1$ Astronomy Department, Faculty of Science, Cairo
University, Giza, Egypt
 \\ email: {\tt wanas@newmail.eun.eg} \\
$^2$Department  of Astronomy , National Research Institute of
Astronomy and Geophysics, Helwan, Egypt.\\email: {\it
 fadymorcos@hotmail.com}\\
$^3$The Egyptian Relativity Group(ERG) URL:www.erg.eg.net
\end{center}

{\bf Abstract}

Solar-Limb Effect is an observational phenomena connected to the
solar gravitational red-shift. It shows a variation of the magnitude
of the gravitational red-shift from the center to the limb of the
solar disc. In the present work an attempt, for interpreting this
phenomena using a general relativistic red-shift formula, is given .
This formula takes into account  the effect of the the Sun's
gravitational field, the effect of the solar rotation, the effect of
inclination of the line of sight and the motion of the observer. In
this study the gravitational field of the Sun is assumed to be given
by Lense-Thirring field instead of the Schwarzschild one. The Earth
is assumed to move along an elliptic orbit. Comparison with a
previous relativistic study and with observation is given.

\section{Introduction}
The solar limb effect is an observational phenomena indicating that
the value of the gravitational red-shift varies from point to point
along the solar disc (cf. Adam (1976, 1979), Peter (1999)). While
the theoretical study using orthodox general relativity (GR) gives
constant value for gravitational red-shift, observations show that
this value increases as we move from the center to the limb of the
Sun's disc. Many authors have attempted to find satisfactory
interpretation for this effect (cf. Mikhail et al.(2002) and
references listed therein).

In a previous attempt (Mikhail et al.(2002)) two of us , in
collaboration with Mikhail have tried to  find an interpretation
using a more general formula for the gravitational red-shift in
the context of GR. In that study the assumptions have been:\\
1) The gravitational field of the Sun is given by the Schwarzschild exterior
solution. \\
2) The observer, on the Earth's surface , moves in a circular orbit
about the Sun.

In the present work we are going to use the same general formula
for the gravitational red-shift, used in the above mentioned study.
The main differences between the present study and above mentioned
one are:\\
(1) The Sun's gravitational field is given by Lense-Thiring
solution of GR in free space.\\
(2) The observer, on the Earth's surface, is assumed to move in an
elliptic trajectory about the Sun.

\section{A General Formula for Red-Shift Variation}
    Kermack, McCrea and Whittaker (1933), established and proved
    two theorems on null-geodesics. As a direct result of
    application of these theorems, they found the following formula:
\begin{equation}
{\h{0}_{1}}=\frac{[\rho_{\mu}~\eta^{\mu}]_{C_1}}{[\bar{\omega}_{\mu}~\eta^{\mu}]_{C_0}}~\lambda_1\,~,
\end{equation}
where we assume that we have two identical atoms at the points
$C_1$ and $C_2$ on the solar equator where there are two
observers $A_1$, $A_2$ respectively, $\lambda_1$,$\h{0}_{1}$ are
the wavelengths of a certain spectral line, as measured by an
observer~$A_1$ at~$C_1$ and B at
  the $C_0$ on the Earth's surface respectively
, $\rho_{\mu}$ gives the components of the tangent to geodesic of
the observer $A_1$ at~$C_1$ and $\eta^{\mu}$ gives the component
of the tangent to the null-geodesic connecting $A_1$ and B and
$\varpi_{\mu}$ gives the components of the tangent to the
trajectory of the observer B.
For the second atom at $C_2$ we can write a formula similar to (2.1)
as:
\begin{equation}
{\h{0}_{2}}=\frac{[\rho_{\mu}~\eta^{\mu}]_{C_2}}{[\bar{\omega}_{\mu}~\eta^{\mu}]_{C_0}}~\lambda_2\,~.
\end{equation}
Now, we can write,
\begin{center}
$\lambda=\lambda_1=\lambda_2$
 \end{center}
as the two atoms are situated on the same great circle (the solar
equator). The quantities between the square brackets in (2.1), (2.2)
are evaluated at the points indicated outside the the brackets,
respectively.\\

As the observer B, on the Earth's surface, measures the
wavelengths coming from the two atoms at $C_1$, $C_2$, he would
expect a difference in the gravitational red-shift given by:
\begin{equation}
 \Delta~Z=\frac{\h{0}_{2}-\h{0}_{1}}{\lambda}.
\end{equation}
Using  (2.1), (2.2),   we can write (2.3) in the form
\begin{equation}
\Delta~Z=\frac{[\rho_{\mu}~\zeta^{\mu}]_{C_2}}{[\bar{\omega}_{\mu}~\zeta^{\mu}]_{C_0}}
-\frac{[\rho_{\mu}~\eta^{\mu}]_{C_1}}{[\bar{\omega}_{\mu}~\eta^{\mu}]_{C_0}}~.
\end{equation}
This formula gives variation of the red-shift of spectral lines
emitted by two identical atoms situated at two different points, on the equator
 of the Sun.\\
In this study we assume that these two atoms are situated in two
symmetric positions relative to the the line connecting the
observer at B and the center of the Sun.
\section{Red-shift Variation in Lense-Thirring Field}

     In this section, we are going to calculate the quantities
     necessary to evaluate the variation of the red-shift given by
     (2.4). For this reason we assume that the exterior
     gravitational field of the Sun, considered as a slowly rotating
     object, is given by the Lense-Thirring metric, (cf. Adler et al (1975))

\begin{equation}
dS^2=\Big(1-\frac{2m}{\varrho}\Big)\,dt^2-\Big(1-\frac{2m}{\varrho}\Big)^{-1}d\varrho^2
-\varrho^2(d\theta^2+\sin^2\theta\,d\phi^2)-\frac{4am}{\varrho}\sin^2\theta
\,dt\,d\phi~,
\end{equation}
where  $m$ is the geometric mass of the
Sun and $(ma)$ is its intrinsic angular momentum. Now, we are going to
use (3.1) to
calculate : \\
1) The tangent of the geodesic, $\rho^{\mu}$, representing the
trajectory of the two similar atoms, assumed to be circular
motion(the
motion is along the solar equator)\\
2) The tangent of the geodesic , $\varpi^\mu$ , representing the
trajectory of the observer B at the Earth's surface (elliptical
motion).\\
3) The component of the null vectors , $\eta^\mu$ , $\zeta^\mu$ ,
tangent to the null-geodesics $\Gamma_1~,~\Gamma_2$ , connecting
$C_1~and~C_0$ ; $C_2~and~C_0$ , respectively. Then substituting
the calculated values into (2.4) we obtain,

\begin{eqnarray}
\Delta~Z&=&\sqrt{\frac{\frac{\gamma^2_{\oplus}}{N^2}
+\frac{\gamma^2_{\oplus}}{r^2}\beta^2-V^2_{\oplus}r^2
-\frac{4m(V^2_{\oplus}a)}{\gamma^2_{\oplus}r}\Big(V^{-1}_{\oplus}+\beta\Big)}
{(\gamma_{\odot}-V^2_{\odot}R^2)+\frac{4am\gamma_\odot}{R^3}\beta}}~\times
\nonumber\\&&\Big[\frac{1+\frac{2am}{R^3}(\beta)}{1+\frac{2am}{r^3}\beta-\frac{2maV_{\oplus}}{r\gamma_{\oplus}}-\sqrt{1-\gamma_{\oplus}\Big(\frac{1}{N^2}
+\frac{1}{r^2}\beta^2\Big)+\frac{4am}{r^3}\beta}}-\nonumber\\&&
\frac{\frac{2am}{R^3}\beta+\Big(1-V_\odot\beta\Big)}
{\Big(1+\frac{2am}{r^3}\beta\Big)\Big(1-V_{\oplus}\beta\Big)+\sqrt{\Big(1-\frac{\gamma_{\oplus}}{r^2}\beta^2+
\frac{4am}{r^3}(\beta)\Big)^2-\frac{\gamma_{\oplus}}{N^2}
\Big(1-\frac{\gamma_{\oplus}}{r^2}\beta^2
+\frac{4am}{r^3}\beta\Big)}} \Big]\nonumber\\&&
\end{eqnarray}
where $\gamma_\odot=1-(2m/r)$,
$\gamma_\oplus=1-(2m/R)$,where (R) and (r) are the radius of the Sun and the mean distance
from the Earth to the Sun respectively, ~$V_{\oplus}$~\& ~$V_{\odot}$~ are  the
orbital angular velocities of the Earth and the atoms on the equator
of the Sun, respectively and $\beta=\frac{\overline{B}}{N}$~where
N \& ~$\overline{B}$ are constants .\\

 It is well known that the angle between any two null geodesics is a
 right angle. So, let us
 consider the angle ~$(\varepsilon)$~ between the projections of the two null geodesics
 (a measurable quantity)  as
 defined by Mikhail et al.(2002) as
\begin{equation}
\cos(\varepsilon)~=a_{ij}~\upsilon^i~\omega^j.
\end{equation}
where~$\upsilon^i$~and $\omega^j$ are  the transport null vectors
along the projection of the  first and second null geodesics,
respectively. It is worth mentioning that the angle $\varepsilon$ is
a small angle. So, it is more convenient to replace it by the angle
$\psi$, between the projection of the radial null-geodesic and the
radius of the Sun passing through the atom. The relation between the
two angles is given by ($\sin\varepsilon=\frac{R}{r}\sin\psi$),
where (R) and (r) are the radius of the Sun and the mean distance
from the Earth to the Sun, respectively. If we neglect terms
containing quantities of the orders $(a/r)^2$ or $(a/R)^2$ and
higher, where $a$ is the angular momentum per unit mass, we get
\newpage
\begin{eqnarray}
\Delta~Z&=&\sqrt{\frac{(\gamma_{\oplus}-V^2_{\oplus}r^2)
-\frac{4amV_{\oplus}}{r}\Big(1+\frac{V_{\oplus}R}{\sqrt{\gamma_{\oplus}}}\sin\psi\Big)}
{(\gamma_{\odot}-V^2_{\odot}R^2)+\frac{4am\gamma_{\odot}}{R^2\sqrt{\gamma_{\oplus}}}\sin\psi}}\times
\nonumber\\&&\Big[\frac{1+\frac{2am}{R^2\sqrt{\gamma_{\oplus}}}\sin\psi}{1+\frac{2amR}{r^3\sqrt{\gamma_{\oplus}}}\sin\psi
-\frac{2amV_{\oplus}}{r\gamma_{\oplus}}-\sqrt{\frac{4amR}{r^3\sqrt{\gamma_{\oplus}}}\sin\psi}}
~-\nonumber\\&&\frac{1+\frac{2am}{R^2\sqrt{\gamma_{\oplus}}}\sin\psi-\frac{2amV_{\odot}}{r\gamma_{\oplus}}-\frac{RV_{\odot}}{\sqrt{\gamma_{\oplus}}}\sin\psi}
{1-\frac{2amV_{\oplus}}{r\gamma_{\oplus}}
-\frac{2amV_{\oplus}R^2}{r^3\gamma_{\oplus}}\sin^2\psi-\sin\psi\Big(\frac{RV_{\oplus}}{\sqrt{\gamma_{\oplus}}}-\frac{2amR}{r^3\sqrt{\gamma_{\oplus}}}\Big)
-\sqrt{\frac{4amR\sin\psi}{r^3\sqrt{\gamma_{\oplus}}}
\Big(1-\frac{R^2}{r^2}\sin^2\psi\Big)}}~\Big].\nonumber\\&&
\end{eqnarray}
Now equation (3.4) represents the difference in the red-shift due to
the theoretical treatment using Lense - Thirring gravitational
field.

\section{Results and Discussion}

Now we are going to evaluate the variation in the red-shift as given by (3.4), in
order to compare it with the well known observational value. We are going to use
the following data for the Sun and the Earth(cf. Arthur
(2000)). These data are summarized in the following table in both
(M.K.S) units and (relativistic units).

\begin{center}
\textbf{Table (1): Dynamical Quantities in M.K.S Units and
Relativistic Units}
\end{center}

\begin{center}
\begin{tabular}{|l|c|c|}
    \hline
  \textbf{Dynamical Quantity }& \textbf{in M.K.S Units} & \textbf{in Relativistic Units}\\ \hline
  $r$~~~~Mean distance form Earth to Sun              & $1.496\times 10^{11}~m$             & 499.0159779 sec \\
  $R$~~~Radius of the Sun               & $6.9599\times 10^{8}~m$             & 2.321591781 sec \\
  $a$~~~~Sun's angular momentum per unit mass               & 273.28 m                          &$ 9.1234\times 10^{-7} sec$\\
  $m$~~~Geometric mass of the Sun                & 1477 m                            &$4.9268\times 10^{-6} sec$\\
  $V_{\odot}$~~The angular velocity of the Sun        & -                                 &$ 2.865\times 10^{-6} rad.sec^{-1}$ \\
  $V_{\oplus}$~~Orbital angular velocity of the Earth        & -                                 & $1.991\times 10^{-7} rad.sec^{-1}$ \\ \hline
\end{tabular}
\end{center}
For the Earth $\gamma_{\oplus}=1$, and for the Sun
$\gamma_{\odot}=0.99999$. By substituting the values of the quantities,
tabulated in Table (1), in equation (3.4), we get
\begin{equation}
\Delta~Z(\psi)=6.1892\times 10^{-6}\times
\Big[\frac{\sin\psi}{1-4.6228\times 10^{-7}\sin\psi}\Big].
\end{equation}
It is clear from this relation that the difference in red-shift on
Solar disc varies as sin the angle $\psi$, which means that there is
a variation in red-shift from  center-to-limb. But this value
contains the variation caused by Doppler shift due to the rotation
of the Sun. So to eliminate this effect (Doppler shift) we consider
two atoms on the equator of the Sun situated at two similar
positions
 $\psi~~and~~-\psi$~then we take the average value.\\
 The \underline{ "Center-to-Limb"} variation in red-shift ($\Lambda$)  is given by the following
equation,
\begin{equation}
\Lambda=\Big(\Delta~Z\Big)_{\psi=90}-~\Big(\Delta~Z\Big)_{\psi=0}
.
\end{equation}
Now by using the numerical values for $\Delta~Z$, given by
equation (4.1) then equation (4.2), will give,
\begin{equation}
\Lambda=2.861\times 10^{-12} \sin^2\psi\,.
\end{equation}
To compare this value with those obtained from the previous study (Mikhail et al(2002)) and from observation, let us calculate the maximum value of (4.3) (for $\psi=90$)in km$/$sec units, we found that
\begin{equation}
\Lambda_{theo}=7.58\times 10^{-7} ~km/sec.
\end{equation}
Although this value is greater than that obtained in the previous study ($
\Lambda=4.8\times 10^{-7}~km$/$sec $), both are still too small compared with the observed value ($\Lambda_{obs}=o.3 ~km/sec$). The ratio between the theoretical and observational values of $\Lambda$ is given by,
\begin{equation}
\frac{\Lambda_{theo}}{\Lambda_{obs}}=2.5\times 10^{-6}.
\end{equation}
So, we still have the same conclusion as in the previous work. The
ratio given by (4.5) indicate that there is some parameter missing in
the theoretical treatment. The order of magnitude of this ratio is
the same as that of the square value of the fine structure constant
($\alpha=\frac{1}{137}$). This gives rise to the idea that the spin
-torsion interaction is the missing parameter, since the coupling
constant of this interaction is the fine structure constant. This
interaction is tested by experiment (Wanas et al.(2000))and by using
observations (Wanas et al.(2002), Sousa and Maluf (2004)).
 A theoretical treatment using the parameterized path equation(Wanas (1998), (2000)), in place of the nullgeodesic one,
 may solve this problem.\\


{{\bf References}}\\
Adam, M.G. (1976) Mon. Not. Roy. Asrton. Soc. \textbf{177}, 687.\\
Adam, M.G. (1979) Mon. Not. Roy. Asrton. Soc. \textbf{188}, 819.\\
Arthur N.Cox. (2000) "\textit{Allen's Astrophysical Quantities}".
$4^{th}$ ed. \\
Adler, R., Bazin, M.,and Schiffer, M. (1975)"\textit{Introduction to General}

{\it Relativity}". $2^{nd}$ ed., McGraw Hill. \\
Kermack, W.O., Mc Crea, W.H. and Wittaker, E.T. (1933) Proc.
Roy. Soc. 

Edin. \textbf{53}, 31.\\
Mikhail, F.I., Wanas, M.I. and Morcos, A.B. (2002) Astrophys. Space
Sci. \textbf{223},

 233; gr-qc/0505132.\\
Peter, H. (1999) Astrophys. J. \textbf{516}, 490.\\
Sousa, A. A. and  Maluf, J. W. (2004) Gen.Rel.Grav. 36, 967-982, gr-qc/0310131v1\\
Wanas, M.I.,(1998), Astrophys. Space Sci., \textbf{258},
 237; gr-qc/9904019.\\
Wanas, M.I.,(2000), Turk. J. of Phys., \textbf{24}, Issue 3,
 473; gr-qc/0010099.\\
Wanas, M.I.,  Melek, M. and  Kahil, M.E. (2000) Grav.Cosmol. 6. \textbf{319}, 322;

gr-qc/9812085.\\
Wanas, M.I.,  Melek, M. and  Kahil, M.E. (2002) Proc. MG IX, Part {\bf B}, 1100;

gr-qc/0306086.\\

\end{document}